\documentclass[aps,prc,10pt,floatfix,amsmath,superscriptaddress]{revtex4-1}
\usepackage{color,graphicx}
\usepackage{dcolumn}
\usepackage{bm}
\usepackage{subfig}
\usepackage{epsfig}
 \usepackage[bookmarksopen]{hyperref}

\def  \f    {\frac}

\def  \bef  {\begin{figure}}
\def  \eef  {\end{figure}}
\def  \be   {\begin{equation}}
\def  \ee   {\end{equation}}
\def  \ba   {\begin{array}}
\def  \ea   {\end{array}}
\def  \bea  {\begin{eqnarray}}
\def  \eea  {\end{eqnarray}}
\def  \beq  {\begin{eqnarray}}
\def  \eeq  {\end{eqnarray}}
\def  \nn   {\nonumber}
\def  \bd   {\begin{displaymath}}
\def  \ed   {\end{displaymath}}
\def  \bse  {\begin{subequations}}
\def  \ese  {\end{subequations}}
\def  \bwt  {\begin{widetext}}
\def  \ewt  {\end{widetext}}

\def  \ba   {{\bf{a_1}}}

\def  \bm   {\bibitem}
\begin{document}
\title{Effect of thermalized charm on heavy quark energy loss}
 \author{Souvik Priyam Adhya}
 \affiliation{High Energy Nuclear and Particle Physics Division, Saha Institute
of Nuclear Physics,
1/AF Bidhannagar, Kolkata-700 064, INDIA}
 \email{souvikpriyam.adhya@saha.ac.in}
 \author{Mahatsab Mandal}
 \affiliation{High Energy Nuclear and Particle Physics Division, Saha Institute
of Nuclear Physics,
1/AF Bidhannagar, Kolkata-700 064, INDIA}
 \email{mahatsab.mandal@saha.ac.in}
 \author{Sreemoyee Sarkar}
\affiliation{Tata Institute of Fundamental Research, Homi Bhabha Road, Mumbai-400005, INDIA}

 \email{sreemoyee.sarkar@tifr.res.in}
 \author{Pradip K. Roy}
 \email{pradipk.roy@saha.ac.in}
 \author{Sukalyan Chattopadhyay}
 \email{sukalyan.chattopadhyay@saha.ac.in}

\affiliation{High Energy Nuclear and Particle Physics Division, Saha Institute
of Nuclear Physics,
1/AF Bidhannagar, Kolkata-700 064, INDIA}

\medskip

\begin{abstract}
The recent experimental results on the flow of $J/\psi$ at LHC show that ample amount
of charm quarks is present in the quark gluon plasma and probably they are thermalized. 
In the current study we investigate the effect of thermalized charm quarks on the heavy 
quark energy loss to leading  order in the QCD coupling  constant. It is seen that the 
energy loss of charm quark increases due to the inclusion of thermal 
charm quarks. Running coupling  has also been implemented to study heavy quark energy 
loss and we find a modest increase in the heavy quark energy loss due to 
heavy-heavy scattering at higher temperature to be realized at LHC energies.
\end{abstract}
\maketitle

\section*{Introduction}
The genesis of a novel state of matter, the quark-gluon plasma (QGP) presumably formed in
the recent experiments at the Relativistic Heavy Ion Collider (RHIC) and 
Large Hadron Collider (LHC) provides us a scope to
investigate the physics of hot and/or dense partonic 
matter \cite{npa757}. One of the excellent probes of QGP, formed in such 
collider experiments, is the quenching of jet  as first anticipated by
Bjorken \cite{Bjorken:1982tu}. The jet quenching or in other words the energy 
loss of the fragmenting partons at high transverse momentum ($p_T$) causes 
depopulation of hadrons in high $p_T$ region.
Most of the calculations of light hadrons ($\pi, \eta$) consider the energy loss due to the 
induced bremsstrahlung radiation and this reproduces the experimentally observed nuclear suppression
 in $Au+Au$ collisions for center of mass
energy $\sqrt{s_{\rm{NN}}}=62-200$ GeV at RHIC \cite{gyulassyreview}.
But recent data from the
PHENIX Collaboration~\cite{raacharm} reveal that 
non-photonic single electron spectrum from heavy meson decays shows larger suppression 
than expected. 
This amount of suppression cannot be explained by radiative energy loss alone because the energy loss of the heavy quark becomes small due to dead cone effect. Thus, in this context, it will be worthwhile to revisit the importance of collisional energy loss of heavy quarks.

The calculation of
collisional energy loss of heavy fermion in Quantum Electrodynamics (QED) or QCD plasma has 
already been studied since past few 
years~\cite{bjorken, abhee05, pradip08, thomaeloss1, thomaeloss2, mgm05, peshier06, djordjevic07, peigne08a,peigne08b, peshier07, sarkar14}.  
Heavy quarks ($Q$) are the important analysing tool of the primordial state of matter as they are
 produced early in the time scale from the initial fusion of partons and do not
influence the bulk properties of the matter. In all the previous estimations mentioned above, 
the presence of thermalized charm quarks in the medium has been neglected. 
 But current experimental evidences from 
ALICE reveal the fact that even heavy quarks along with the light quarks ($q$) and gluons ($g$) could be a possible composition of the medium and their 
impact on different physical quantities can not be ruled out.
\\
 Recently the ALICE collaboration has reported a value of $0.58\pm 0.01(stat)\pm0.09(syst)$ for the Nuclear Modification Factor ($R_{AA}$) of $J/\psi$ measured for $Pb-Pb$ collisions at $\sqrt{s_{NN}}=2.76$ TeV using the Muon spectrometer ($2.5<y<4$)\cite{jpsiplb}. However, at low transverse momentum ($p_T< 4 $ GeV), significantly larger values of $R_{AA}$ were measured compared to the measurements at lower energies in RHIC. This observation indicates a substantial contribution to the $J/\psi$ production from charm quark recombination. The same data also indicates a non-zero elliptic flow coefficient ($v_2$) with a largest value of $0.116\pm0.046(stat)\pm0.029(syst)$ for $J/\psi$ in the transverse momentum range $2\leq p_T< 4$ GeV/c \cite{abbas13}. These two observations seem to suggest a significant presence of thermalized charm quarks in the deconfined partonic matter produced in the $Pb-Pb$ collisions.
 Consequently these findings demand further modification of formalism of heavy quark energy 
loss with thermalized $c\bar{c}$ 
pairs taken into consideration. We, in the current study develop a complete theoretical 
framework to evaluate the energy loss of an energetic heavy quark including all 
possible scatterings with the medium particles 
to the leading order in the QCD coupling constant ($\alpha_s$).

In all the previous studies 
it has been established that in the medium the heavy flavor energy loss is plagued with infrared
divergences. In the non-relativistic plasma, this can be removed 
with the help of the Coulomb or electric interaction. However, 
the problem becomes nontrivial with the relativistic
plasma where both the
electric and the magnetic interactions are important. To circumvent the problem, the usual way is to 
implement Braaten and Yuan's prescription \cite{yuan91}. In this prescription, one separates the 
integration into two domains with an arbitrary
cut off momentum scale $q^* (gT\ll q^* \ll T)$ where $g^2=4\pi \alpha_s$ and $T$ is defined as the temperature of the partonic medium. In the domain of exchange of hard gluons, 
(where, momentum ($q$) transfer $q\sim T$) one uses a tree level propagator and in the 
region of soft transfer (where $q\sim gT$) 
hard thermal loop (HTL) corrected propagator is required 
to provide the necessary screening from the infra-red (IR) divergence at the Debye scale
$gT$ \cite{pisarski90,braaten91}.  It is important to note that both the contributions from the hard as well
as the soft sectors must be added together to cancel the intermediate
momentum scale making the final result independent of this arbitrary cut-off parameter ~\cite{thomaeloss2}. 
In the current paper, we adopt the method of Braaten and Yuan to compute
 heavy quark energy loss. 

 We also incorporate the effects of running coupling ($\alpha_{eff}$) in the calculation. In QCD medium, effective 
 field theory imposes running of the coupling constant instead of a fixed one at very high 
temperature ($T\gg \Lambda_{s}$). Implementation of running coupling has already been used to compute
 various physical quantities since past few years \cite{uphoff11, gossiaux08, peigne08b,lusaka11,tolos13}. Here, we employ 
 the parametrization of $\alpha_{eff}$ in the time-like sector from Ref.\cite{dokshitzer96} by extending it to the 
the space-like sector \cite{dokshitzer96, gossiaux08, uphoff11}. This is a non-perturbative technique where $\alpha_{eff}$ 
remains finite even in the infrared domain \cite{dokshitzer96}. The computation of heavy quark energy loss with thermalized  $c\bar{c}$ in the medium using
 $\alpha_{eff}$ is another new component of the current manuscript. 
 These improvisations aim at studying the experimental values of $R_{AA}(p_T)$ and $v_2(p_T)$ complementing 
 the theoretical predictions.
 

%
\section{Collisional energy loss}

The motion of a heavy quark in partonic matter looks similar to that of a test particle
in the plasma. Hence, its motion can be treated as a typical Brownian
motion problem.
\begin{figure}[htb]
\centering
   \begin{tabular}
   {c@{\hspace*{0.25mm}}c@{\hspace*{0.25mm}}c@{\hspace*{0.25mm}}c@{\hspace*{0.25mm}}c}
\resizebox{3.5cm}{2.0cm}{\includegraphics{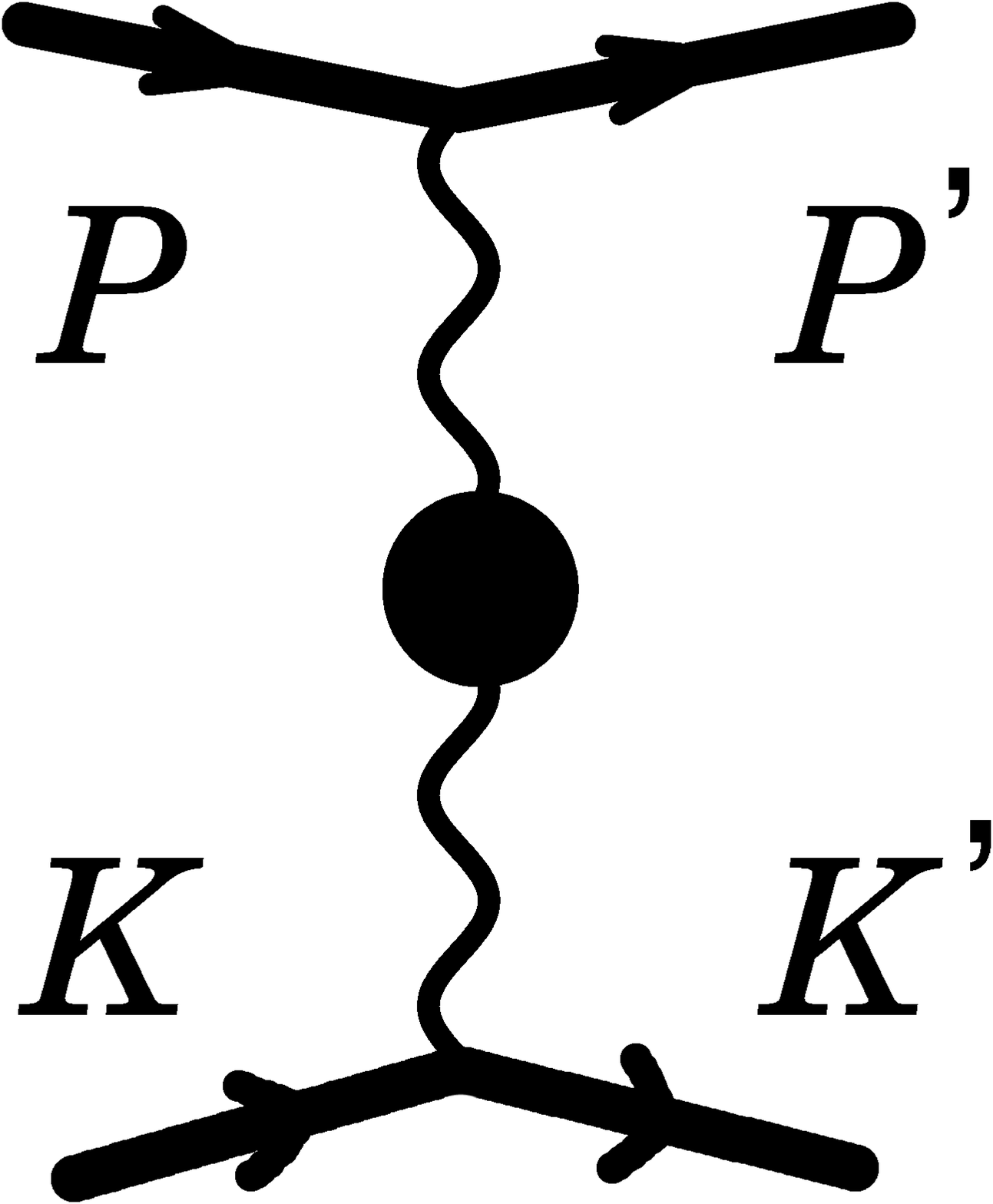}}

&
\resizebox{3.5cm}{2.0cm}{\includegraphics{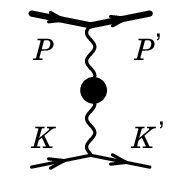}}

&
\resizebox{3.5cm}{2.0cm}{\includegraphics{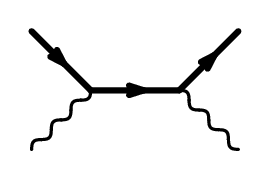}}
   &
    \resizebox{3.5cm}{2.0cm}{\includegraphics{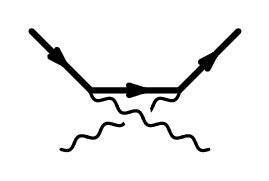}}
   &
\resizebox{3.5cm}{2.0cm}{\includegraphics{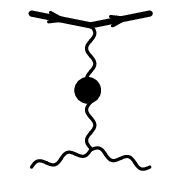}}
   \\
     (a) & (b) & (c) & (d) & (e)

  \end{tabular}
   \caption{Amplitudes for heavy quark elastic scattering in a QCD plasma. Fig.(a) 
   corresponds to heavy quark-heavy quark scattering. Diagrams (b), (c), (d) and (e) correspond to $Qq$ and $Qg$ scatterings in $s, t$ and $u$ channels. The blob in (a), (b) and (e) 
   denotes the resummed hard thermal loop boson propagator, which is
 required to screen the $t$-channel contribution in the infrared domain.}
   \label{fig1}
   \end{figure} 
The kinematics of the test particle in the plasma can be described by Boltzmann equation. 
In this current scenario, since there are no 
external forces and the system is homogeneous, Boltzmann 
equation reduces to,
\bea
\frac{\partial f_p}{\partial t}=-\mathcal{ C}[f_p],
\label{boltz_2}
\eea
where the right hand side of the above equation represents the collision integral. 
In the present paper, we consider a high energy heavy ($Q$) quark of mass 
$m_1$, energy $E_p$ and momentum $p$ propagating through a
plasma consisting light quarks ($q$), gluons ($g$) and charm quark in equilibrium at a 
temperature $T$. The injected heavy quark has 
small fluctuating part ($\delta f_p$, $f_p^0\gg \delta f_p$) which vanishes suffering 
collisions with other particles of the medium.
In the present work, we only consider $2\rightarrow 2$ ($P+K\rightarrow P'+K'$)
scattering processes. 
The explicit form of the collision integral then becomes,
\bea
\mathcal{C}[f_p]&=&\frac{1}{2E_p}\int \frac{d^3k}{(2\pi)^3 2E_k}
\frac{d^3p^{'}}{(2\pi)^3 2E_p'}\frac{d^3k^{'}}{(2\pi)^3 2E_k'}
 \delta f_p[PSF]\nn\\
&\times&(2\pi)^4 \delta^4(P+K-P'-K')
\frac{1}{2}\sum_{spin}{\cal| M |}^2,
\label{collision_term}
\eea
where, [PSF] denotes the phase space factor. ${\cal| M |}^2$ in the above 
equation contains the information about the interaction concerned. 
In a $P+K\rightarrow P'+K'$ scattering process, the energy and momentum 
variables are denoted as $X=(E_X, \vec X)$, (where $X=P,K,P',K'$). 
The thermal phase space contains the information of elastic 
scattering processes (Fig.(\ref{fig1})) and its form changes with the nature of the scatterings. The details of phase-space factor for different processes
 will be discussed later in the text. 
 
 In the relaxation
time approximation, Eq.(\ref{boltz_2}) can be expressed as,
\bea
\frac{\partial \delta f_p}{\partial t}=-\mathcal{ C}[ f_p]=-\delta f_p \Gamma(p).
\label{boltz_3}
\eea
where $\Gamma(p)$ is the particle interaction rate, given by
\bea
\Gamma(p)&=&\frac{1}{2E_p}\int \frac{d^3k}{(2\pi)^3 2k}\frac{d^3p^{'}}{(2\pi)^3 2E_p'}\frac{d^3k^{'}}{(2\pi)^3 2k'}
 [PSF](2\pi)^4 \delta^4(P+K-P'-K')
\frac{1}{2}\sum_{spin}{\cal| M |}^2.
\label{damprate_term}
\eea
The heavy quark energy loss ($-dE/dx$) is then obtained by averaging over the interaction
rate times the energy exchange per scattering and dividing
by the velocity of the injected heavy quark,
\bea
\frac{dE}{dx}=\frac{1}{v_p}\int d\Gamma \omega.
\label{e_loss_def}
\eea
The factor $\omega$ in Eq.(\ref{e_loss_def}) is essential in making $dE/dx$ finite 
within resummation perturbation theory. The calculation of the
interaction rate mentioned in Eq.(\ref{damprate_term}), using tree level diagram, 
suffers quadratic infrared divergence. But in case of energy loss 
the extra $\omega$ factor softens the infrared divergence to only logarithmic. 
On the other hand, the resummed effective boson propagator makes $\Gamma$ to be only logarithmically
divergent which in turn makes $-dE/dx$ finite \cite{thomaeloss1,thomaeloss2}. 
The energy loss of a heavy-quark propagating through a hot quark-gluon plasma 
can be calculated either from the field theory approach or using effective kinetic theory. 
In the present paper we follow the latter one. 

In order to estimate the charm quark energy loss it is necessary 
to have an idea of number of scattering 
centers present in the bath. It is well known that $\Gamma=n\sigma v$, where $n$ is the density of the plasma
particles, $\sigma$ is the collision cross section and $v$ is the velocity of the particle,
which is equal to the velocity of light in case of relativistic particles \cite{blaizotphysrept99}. 
The number 
density of charm quarks, light quarks
and gluons present in the medium has been estimated by the following formula \cite{braun00},
\bea
n_i=\frac{g_i}{(2\pi)^3}\int_0^{\infty}\frac{ d^3p }{\rm{e}^{E_i/T}\pm 1}
\eea
where, $g_i$ is the degeneracy factor, $E_i$ is the total energy of the particles ($i=Q,q,g$). 
In Table.(I)  the variation of the number densities 
with the temperature of the medium can be observed. From the table it is evident that at temperatures 
relevant to LHC energies heavy quark density is quite significant along 
with light quarks and gluons. 
Hence, to construct a consistent formalism of heavy quark energy loss it is indeed necessary 
to incorporate the scatterings of injected parton with heavy quarks present in the medium. 
In the following section, we compute the contribution of heavy 
quark scatterings to the total energy loss.
\begin{table}
\label{table_1}
\begin{center}
\begin{tabular}{|c|c|c|c|c|}\hline\hline
{Temperature (GeV)}&{$n_q+n_{\bar q}(fm^{-3})$} &
{$n_g (fm^{-3})$}&
{$n_Q+n_{\bar Q}(fm^{-3})$}\\\hline
0.3 & 7.70374 & 6.8478 &0.5795  \\ \hline
0.4 & 18.2607& 16.2317& 2.5511\\ \hline
 0.5&
35.6655& 31.7026& 7.3500\\ \hline
0.6&
61.6299& 54.7822&  16.0630\\ \hline

\end{tabular}
\end{center}
\caption{Variation of number density of heavy quarks, light quarks and gluons with temperature}
\end{table}

 \subsection{Contribution of $QQ\rightarrow QQ$ scatterings}
In this section, we illustrate the energy loss where the heavy 
quark ($p$) interacts with the thermal  heavy quarks of momentum $k$.
Following Eq.(\ref{e_loss_def}) the explicit form of heavy quark energy loss
turns out to be, 
\bea
\label{lossdef}
\Big(-\f{dE}{dx}\Big)_{QQ\rightarrow QQ} =\frac{1}{v_pE_p} \int_{p'}\int_k  
\int_{k'} (2\pi)^4 \delta^4(P+K-P'-K') \f{1}{2} \sum_{\rm spins} |{\cal M}|^2 
\omega  [PSF]  ,
\label{orieqn}
\eea
where, $\omega =(E_p -E_{p'})$ is the exchanged energy in an elastic scattering and $\int_{p'}$ 
denotes $d^3p'/(2E_{p'}(2\pi)^3)$. The above expression has been obtained by inserting the factor 
$\omega$ in Eq.(\ref{damprate_term}) and multiplying by a factor of $2$ to consider both the quark 
and the anti-quark scatterings ($Q\bar Q\rightarrow Q\bar Q$). 
Eqn.(\ref{orieqn}) is a general expression and is true for all kind of scatterings suffered by the heavy quark with the particles in the medium. The $t-channel$ contribution of the matrix amplitude gives a contribution proportional to $\int dq/q^3$ to the Eqn.(\ref{orieqn}). This factor multiplied with the energy transfer $(-q\leq \omega \leq q)$ softens the divergence to a logarithmic one. Further, self energy appearing on the exchange line cuts off this logarithmic divergence to give a term $\sim \rm{ln}(q_{max}/m_D)$ where $q_{max}(\sim \sqrt{E_p T})$ is the maximum momentum transfer per scattering and $m_D(\sim gT)$ is the Debye mass. Thus the logarithmic contribution is $\sim \rm{ln}(1/g)>> 1$ for $g<<1$. This kind of leading log treatment extracts the coefficient of the logarithmic divergence which we have followed in our calculations to give results upto leading logarithmic accuracy. It might be mentioned here that the energy loss receives dominant contribution when the transferred momentum is 
assumed to be soft $(0< q<m_D)$ and $q<< T, p, k$ which is precisely our kinematical domain for 
leading logarithmic calculation \cite{arnold00,arnold03}.

 In case of
 heavy and light quark interaction, the possibility of back scatterings in phase space factor ($[PSF]$) 
can be excluded. However, for the collision between heavy quarks one has to retain this process along with the forward one. This $[PSF]$ factor is then given by \cite{abhee05},
\bea
[PSF]=&&f_{E_k}(1-f_{E_{k'}})(1-f_{E_{p'}}) + (1-f_{E_k}) f_{E_{k'}}f_{E_{p'}}\nn\\ 
\eea
where the first part of the above expression corresponds to forward scattering and the later part is included for back scattering contribution. Thus, we have,
\bea 
[PSF]=&&(f_{E_k}-f_{E_{k'}})\left [ 1 + \bar f_{q_0} - f_{E_{p'}} \right ]\nn\\
&&\simeq -\frac{df_{E_k}}{dE_k}q_0 \left[\frac{T}{q_0}+\frac{1}{2} \right ],
\eea
where, $\bar f_{q_0}$ is the boson distribution function and is given by $\bar f_{q_0}=(\exp(q_0/T)-1)^{-1}$ 
and $q_0=\omega$.  

The diagrams required for the calculation of the heavy quark energy loss are given in Fig.(\ref{fig1}). It is to be noted that all these diagrams contribute at leading log order.  Among these set of diagrams, the first one has not been calculated yet which is the main focus of this present paper. Apart from these diagrams, there is contribution from the $s-channel$ scattering process which infact contribute at the same order in $\alpha_s$ as the $t-channel$ one. Thus, for consistency, we have outlined the calculation for the $s-channel$ in the Appendix. However, the contribution from the $s-channel$ is suppressed in comparison to the $t-channel$ which will be discussed in detail in the Results section. 

In this section we present the derivation of heavy quark energy loss in 
the soft regime defined earlier as $|q|\ll|q^\star|$ for the $t-channel$ process.  The soft contribution to the 
energy loss is evaluated in the region of phase space where the exchanged gluon has 
momentum of the order of $gT$. In this kinematical region, the gluon propagator has to be 
modified using the HTL resummation method in order to incorporate the in-medium 
effects.

In Eq.(\ref{orieqn}), the 3-momentum delta function is used to eliminate the $p^{\prime}$
integral,
\bea
\int d^3p^{\prime}\delta ^{(3)}(p+k-p^{\prime}-k^{\prime})=1.
\eea
Next, we introduce an energy variable $\omega$ in the integrand of Eq.(\ref{orieqn}), 
\bea
\delta(E_p+E_k-E_{p^{\prime}}-E_{k^{\prime}})=\int^{\infty}_{-\infty}\delta(E_p-E_{p^
{\prime}}-\omega)\delta(\omega-E_{k^{\prime}}+E_k)d\omega.
\eea
Two delta functions of the energy variable can then be written as,
\bea
\delta(E_p-E_{p^{\prime}}-\omega)&=&\frac{E_{p^{\prime}}}{E_p}\delta \left(\omega-\vec
v_p.\vec q-\f{t}{2E_p}\right)\nn\\
\delta(\omega-E_{k^{\prime}}+E_k)&=&\frac{E_{k^{\prime}}}{E_k}\delta \left(\omega-\vec v_k.\vec
q+\f{t}{2E_k}\right),
\eea
where, $\vec q=\vec p-\vec p^{\prime}=\vec k^{\prime}-\vec k$ and $\vec v_k=\vec k/E_k$. 
From this point it is convenient to change the integration
 variables from $k$, $k'$ to $k$ and $q$ respectively.
Using the expressions of the $\delta$ functions given above, the energy loss 
of the heavy quark given in Eq.(\ref{orieqn}) is written as,
\bea
\left(-\f{dE}{dx}\right)\Big|^{\rm soft}_{QQ\rightarrow QQ}&=&\f{(2\pi)^4}{(2\pi)^9 E_p2^3.2v_p}\int d^3k
\int d^3q\int^{\infty}_{-\infty} d\omega 
\Big[\f{E_p^\prime E_k^\prime}{E_p^\prime E_k E_k^\prime E_p E_k}\Big][PSF]\nn\\
&&\delta
\Big(\omega-\vec v_p.\vec q-\f{t}{2E_p}\Big)\delta \Big(\omega-\vec v_k.\vec
q+\f{t}{2E_k}\Big)\omega \mid {\cal M}\mid^2 .
\label{hQ_eloss}
\eea
In order to proceed further, it is necessary to introduce the 
form of the interaction. In the Coulomb gauge the matrix element ${\cal M}$ for 
the $QQ\rightarrow QQ$ process can be expressed as follows \cite{thomaeloss1},
\bea
{\cal M}= g^2 
D_{\mu \nu } (q) \bar{u} (p^\prime, s^\prime) \gamma^\mu u (p, s) 
\bar{u} (k^\prime, \lambda^\prime) \gamma^\nu u (k, \lambda),
\eea
where $\alpha_s$ is the strong
coupling constant.
In the above mentioned gauge, only non-vanishing
 components of the bosonic propagator are,
\bea
 \Delta^{00}(Q)&=&\Delta_{L}(q_0,q),\nn\\
\Delta^{ij}(Q)&=&\Delta_{T}(\delta^{ij}-\hat {q^i}\hat{q^j}).
\label{non-zero_prop}
\eea
$\Delta_{L}$ and $\Delta_{T}$ are the longitudinal and transverse components of the boson 
propagator and are given by \cite{bellacbook},
 \bea
\Delta_L(Q)&=&\f{-1}{q^2+m_D^2(1-\f{x}{2}\log(\f{x+1}{x-1}))}\nn\\
\Delta_T(Q)&=&\f{-1}{q_0^2-q^2-\f{m_D^2}{2}\Big(x^2+\f{x(1-x^2)}{2}\log(\f{x+1}{x-1})\Big)},
\eea
 where, $m_D$ is the Debye mass and $x=\omega/q$.
 
Following Eq.(\ref{non-zero_prop}) the non-zero components of the matrix element are~\cite{djordjevic06},
\bea
{\cal M} &=& g^2 \Delta_L (q) \bar{u} (p^\prime, s^\prime) \gamma^0 u (p, s) 
\bar{u} (k^\prime, \lambda^\prime) \gamma^0 u (k, \lambda) \nonumber \\
&+& g^2 \Delta_T (q) (\delta^{i j}-\hat{q}^i \hat{q}^j)
\bar{u} (p^\prime, s^\prime) \gamma^i u (p, s) 
\bar{u} (k^\prime, \lambda^\prime) \gamma^j u (k, \lambda).
\eea 
%

The matrix element given above has to be squared, averaged over 
initial spin $s$ of the jet and summed over final spins. After evaluating 
the Dirac traces, we obtain,
\bea
\f{1}{2}\sum_{spins}\mid {\cal M}\mid^2&=& 32 g^4 E_p^2\Big\{\mid
\Delta_L(Q)\mid^2 E_k^2+2E_k\mid\vec k \mid [{(\vec v_p.\hat k)-(\hat q.\hat
k)(\hat q.\vec v_p)}]\mbox{Re} [\Delta_L(Q)\Delta_T(Q)^{*}]\nn\\
&+&\mid \Delta_T(Q)\mid^2\mid\vec k \mid^2[(\vec v_p.\hat k)-(\hat q.\hat
k)(\hat q.\vec v_p)]^2\Big\}.
\eea
While, writing the above equation it has been assumed that leading logarithmic order contributions come from the 
region where the exchanged energy $\omega$ is small. Hence, $E_p'\simeq E_p$ and $E_k'\simeq
E_k$. 
The soft contribution to the heavy quark energy loss then reduces to,
\bea
\left(-\f{dE}{dx}\right)\Big|^{soft}_{QQ\rightarrow QQ}&=&\f{ g^4 C_F}{4v_p^2\pi^3 }\int  dq\int k
dk\f{1}{E_k}\int_{-v_p q}^{v_p
q}\omega^2\left(-\right)n_F'(E_k)d\omega\nn\\
&\times&\Big\{\mid \Delta_L(Q)\mid^2\f{E_k^2}{v_k }+\mid
\Delta_T(Q)\mid^2k^2\f{1}{2v_k }\Big[1-\f{\omega^2}{(v_k
q)^2}\Big]\Big[v_p^2-\f{\omega^2}{(v_k q)^2}\Big]\Big\}.
\label{dedxsoft1} 
\eea
The above equation has been derived by using the following 
integrals over the angles of $k$, 
\bea
&&\int\f{d\Omega_k}{4\pi}\delta(\omega-\vec v_k.\vec q)=\f{1}{2v_k q};\nn\\
&&\int\f{d\Omega_k}{4\pi}\delta(\omega-\vec v_k.\vec q)\Big[\vec v_p.\hat
k-\f{\omega}{v_k q}(v_k \hat q).\hat k\Big]=0;\nn\\
&&\int\f{d\Omega_k}{4\pi}\delta(\omega-\vec v_k.\vec q)\Big[\vec v_p.\hat
k-\f{\omega}{v_k q}(v_k \hat q).\hat k\Big]^2=\f{1}{4 v_k
q}\Big[1-\f{\omega^2}{(v_k q)^2}\Big]\Big[v_p^2-\f{\omega^2}{(v_k q)^2}\Big].
\eea

Further evaluation of $(-dE/dx)$ in Eq.(\ref{dedxsoft1}) cannot be performed analytically and has to be
solved numerically.

It is to be noted that contribution from the integration domain where 
screening effect is absent can be extracted from Eq.(\ref{dedxsoft1}) 
by putting screening mass to zero in the denominator of the propagator~\cite{bellac97}.
Final result of heavy quark energy loss scattering off thermalized charms in the 
medium is obtained by adding both the  soft and hard contributions,
 \bea
 -\f{dE}{dx}\Bigg|_{QQ\rightarrow QQ}=-\f{dE}{dx}\Bigg|^{soft}_{QQ\rightarrow QQ}+
 -\f{dE}{dx}\Bigg|^{hard}_{QQ\rightarrow QQ}
 \label{tot_heavy_heavy}
 \eea
 In this regard it would be worthwhile to mention the contribution of scatterings of 
 heavy quark with other particles of the medium. The  results for scatterings with 
 light quarks can be obtained from Eq.(\ref{tot_heavy_heavy}) in the limit 
 $v_k\rightarrow 1$ and the result matches with the findings of Ref.\cite{thomaeloss2}. 
 The contribution of $Qg\rightarrow Qg$ scatterings can be read from Eqs.(1), (6) and (7) 
 of Ref.\cite{thomaeloss2}.  Hence, the complete expression for heavy quark energy 
 loss can be obtained by adding all possible scatterings mentioned in 
 Eq.(\ref{tot_heavy_heavy}) and in Ref.\cite{thomaeloss2}.

\subsection{Implementation of running coupling}
In the above calculations, we have explicitly assumed that the value of QCD coupling
 $\alpha_s$ is kept fixed. However, an effective calculation of the cross-sections for the process $QQ(\bar Q)\rightarrow QQ(\bar Q)$ and 
 estimation of the energy loss of the heavy quark can be performed taking into account the running coupling. In this work, 
 we incorporate the effective coupling on all order resummations of perturbation theory including all non-perturbative effects as observed 
 in Ref.\cite{dokshitzer96,gossiaux08,uphoff11} to calculate the energy loss. The transition from $\alpha_s$ to the running regime ($\alpha_{eff}$) is stated as \cite{dokshitzer96,gossiaux08,uphoff11},
\bea
\alpha_s \rightarrow  \alpha_{eff}(Q^2)\nn\\
\eea
where,
\begin{align}
\label{alpha_s_continued}
 \alpha_{eff}(Q^2)= \frac{4\pi}{\beta_0} \begin{cases}
  L_-^{-1}  & Q^2 < 0\\
  \frac12 - \pi^{-1} {\rm arctan}( L_+/\pi ) &  Q^2 > 0
\end{cases}
\end{align}
with $Q^2=\omega^2-q^2$, $\beta_0 = 11-\frac23\, n_f$  and
$L_\pm = \ln(\pm Q^2/\Lambda^2)$ with $\Lambda=0.263 \,$GeV.
Moreover, we have also treated Debye mass ($m_D$) to be a function of both $Q^2$ and $T$, \rm{i.e.},
\bea
m_{D}^2\equiv m_{D}^2(T, Q^2)= 4\pi \alpha_{eff}(Q^2) (1+n_f/6) T^2
\eea
The results for the energy loss of the heavy quark with the inclusion of running coupling are presented in the following section.
\section{Results}
 \begin{figure}
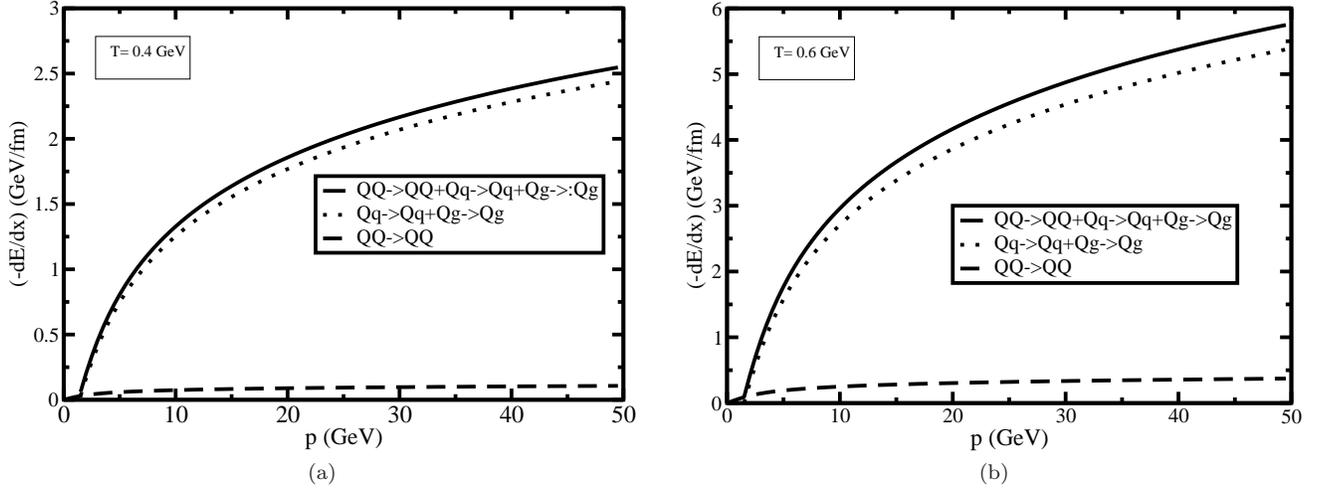

\centering
     \subfloat[][]
  {\includegraphics[height=6cm, angle=0]{latest_.4.eps}\label{<figure1>}}~~~~~
 \subfloat[][]{\includegraphics[height=6cm, angle=0]{latest_.6.eps}\label{<figure2>}}
  \caption{Energy loss $dE/dx$ of a charm quark as a function of its momentum for
 $T=0.4$ GeV (a) and $T=0.6$ GeV (b).}
 \label{fig2}
  \end{figure}
 In the present paper we calculate the energy loss of a heavy quark in a medium where, 
in addition to the light quarks and gluons, thermalized heavy quarks are also present.
The heavy quark loses energy in the hot medium {\em via} all $t$, $s$ and $u$ channel processes. 
In Fig.(\ref{fig2}) the total heavy quark energy loss due to scatterings with medium particles 
has been compared with  the known result of light quarks and gluons scatterings \cite{thomaeloss2}. 
We also plot the contribution of heavy quark scatterings off thermalized heavy quarks in the medium. 
In this calculation, the momentum has been scaled to an upper limit of 
$q_{max} = \sqrt{4E_pT}$ in compliance with the results in~\cite{thomaeloss1}.  
With our present calculation, we present plots of 
$-dE/dx$ with the momentum, considering the temperatures of 400 MeV and 600 MeV
respectively relevant to the plasma temperature produced at LHC.
For the plots the heavy quark mass has taken to be $1.25$ GeV and strong 
coupling constant $\alpha_s = 0.3$. 

From the two plots in Fig.(\ref{fig2}) it is evident that the contribution to the energy loss due 
to $QQ \rightarrow QQ$ scatterings increases with temperature like other two contributions. 
The observation is consistent with the earlier findings of number densities in the current paper. 
Increase in temperature increases number densities which in turn increases interaction rate 
as well as particle energy loss.
It is observed that by including the process
$QQ \rightarrow QQ$ the total heavy quark energy loss increases by $~5 \%$ and $~8 \%$ for temperatures
$400$ MeV and $600$ MeV respectively at a momentum of $25$ GeV.
\begin{figure}
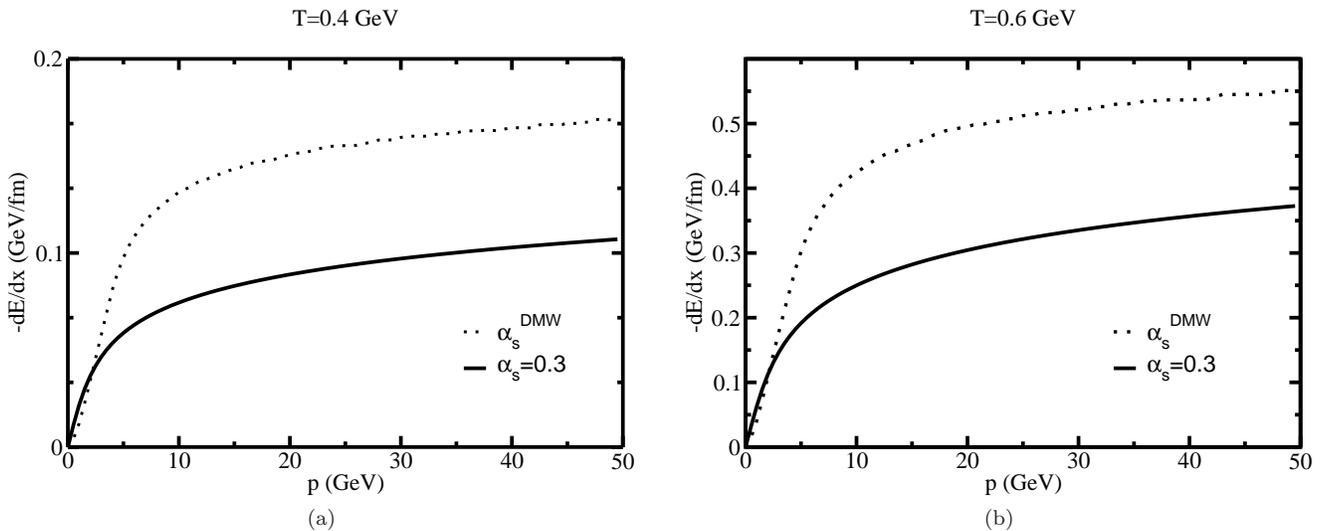

     \centering
     \subfloat[][]{\includegraphics[height=6.5cm, angle=0]{run_dok_.4.eps}\label{<figure1>}}~~~~~
     \subfloat[][]{\includegraphics[height=6.5cm, angle=0]{run_dok_.6.eps}\label{<figure2>}}
     \caption{Comparison of running coupling ($\alpha_{eff}$) on energy loss for 
 $QQ \rightarrow QQ$ scatterings at $0.4$ GeV (a) and at $0.6$ GeV (b) with the fixed value of the coupling constant ($\alpha_s$).}
     \label{fig4}
\end{figure}
We have assumed a fixed value of the coupling constant for estimation of these results.
%
Now, using the expression of $\alpha_{eff}$ as in Eq.(\ref{alpha_s_continued}), we  evaluate the energy loss with charm quarks as particles of the medium. In the left panel of Fig.~(\ref{fig4}), a comparative study has been presented of the charm quark 
energy loss due to scattering with another heavy quark 
with constant and running coupling for a temperature of 0.4 GeV.
In the right panel, similar comparisons have been
 performed for a fixed temperature of 0.6 GeV. 
In both of these panels, we find that the energy loss of the heavy quark under the 
inclusion of the t-channel $QQ\rightarrow QQ$ scattering,
increases with temperatures as already obtained in the constant $\alpha_s$ case. 
We also find that the energy loss is greater when the coupling constant 
is taken to be running (Eq. (\ref{alpha_s_continued})) as compared to the fixed one.
These findings are consistent
 with the results obtained in Ref.\cite{peigne08b, gossiaux08, uphoff11}.
 In addition, we also calculate 
 In Fig.\ref{fig5}, we have shown the contribution of the scattering process in the the $s-channel$ $(Q\bar{Q}\rightarrow g \rightarrow Q\bar{Q})$. A brief calculation of this process has been given in the Appendix. For a typical momentum of $25$Gev, we find that the $s-channel$ contribution is $\sim 0.2\%$ (for both T=0.4 GeV and 0.6 GeV) to the total energy loss of the heavy quark with $s-channel$ and $t-channel$ combined. We reconfirm that the relative contribution of the $s-channel$ is significantly lower than the corresponding $t-channel$ processes that we have considered for our calculation of the energy loss of the heavy quark and hence can safely be ignored\cite{arnold00, arnold03}.


\section{Summary and Conclusions}
In the present study, we have discussed the theory of heavy quark energy loss in a hot QCD plasma with specific emphasis on the impact of 
thermalized charms on the heavy flavor energy loss. In all the previous calculations of 
charm quark energy loss it has been assumed that the thermalized partonic medium is devoid
 of heavy quarks and consists of only light quarks and gluons. 
 But the recent ALICE data has shown non-zero elliptic flow for $J/\psi$ which is possible only if 
 charm quarks gets thermalized 
in the medium. 
Hence, we have elucidated a consistent formalism of heavy quark energy loss, where
 all the possible scatterings with the particles and antiparticles of the medium 
 ($Qq \rightarrow Qq, Qg \rightarrow Qg$ and $ QQ \rightarrow QQ$) have been taken into consideration. 
From the numerical 
 studies presented in the paper it is observed that the $QQ \rightarrow QQ$ scattering has a modest
 effect on the charm quark energy loss. 
It has also been observed that with the 
 increase in temperature heavy quark energy loss due to $QQ \rightarrow QQ$ scattering  increases thereby enhancing the total energy loss.
In addition, the heavy quark energy loss increases fairly under inclusion of running coupling.
 These observations are consistent with the nature of number densities of plasma particles with temperature. 
 In fact, this mechanism proves to be
quite an efficient one for the energy transfer into the plasma 
which might have possible implications in the explanation of the regeneration of 
$J/\psi$ at LHC. 
From the observables point of view we know that final heavy quarks are tagged in the heavy ion collision experiments
to measure heavy 
flavour elliptic flow and nuclear modification factor. Now, in these experiments another possible
source of heavy quarks could be scattering of high $p_T$ gluons with thermalized charm quarks which contribute to
 produce high $p_T$ D-mesons. 
  Introduction of this process in the present calculation is a delicate issue. It is also very difficult to disentangle the high $p_T$ heavy quark jet scattered from a thermal heavy quark and high $p_T$ heavy quark jet produced from $g Q_{Th} \rightarrow g Q$. Thus, we do not address this possibility in the present work. In this paper our main concern has been to develope 
consistent theoretical formalism of heavy quark energy loss with all possible scatterings. 
Effects of current findings on different physical observables at LHC will be reported elsewhere.
 \bigskip
\begin{figure}
     \centering
     \includegraphics[height=6.5cm, angle=0]{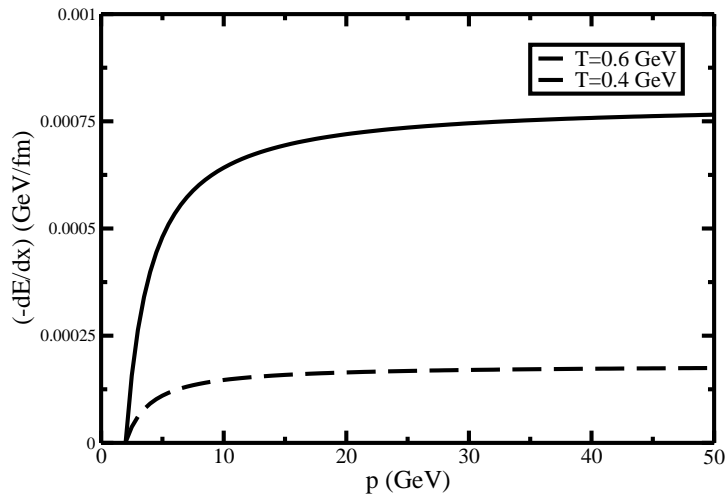}
     \caption{$s-channel$ contribution for the process 
 $Q\bar{Q} \rightarrow Q\bar{Q}$ scatterings at $0.4$ GeV and at $0.6$ GeV  with the fixed value of the coupling constant ($\alpha_s$).}
     \label{fig5}
\end{figure}
\section*{Acknowledgment}
One of the authors (S. P. A.) would
like to thank UGC, India (Serial No. 2120951147) for
providing the fellowship during the tenure of this work. S. P. A. would also like to thank Prof. Jan-e-Alam for fruitful discussions regarding different aspects of the paper. 
\section{Appendix}
We briefly outline the calculation of the $s-channel$ scattering process. Following the steps in \cite{peigne08a}, Eq.\ref{orieqn} is modified for $s-channel$ as,
\bea
\Big(-\f{dE}{dx}\Big)_{Q\bar{Q}\rightarrow Q\bar{Q} (s-channel)}=\f{1}{v_p}\int\f{d^3 k [PSF]}{(2\pi)^32 E_k}\int^{4m^2-s}_{0}du(4m^2-s-u)(F_1)(F_2)\mid{\cal M}\mid^2 
\label{schaeqn}
\eea
where the functions are evaluated as,
\bea
F_1=\Big(1-\frac{s-2\left.m^4\right/t}{s\left(1+\left.m^2\right/t\right)-2\left.m^4\right/t}\frac{E_k}{E_p}\Big)\nn\\
F_2=\left(\frac{s\left(1+\left.m^2\right/t\right)-2\left.m^4\right/t}{\left(s\left(s-4 m^2\right)\right)^{3/2}}\right)
\eea
and the matrix amplitude squared is given for the $s-channel$ as\cite{halzenbook},
\bea
\mid{\cal M}\mid^2=\f{64\pi^2\alpha_s^2}{9}\f{(2m^2-u)^2+(2m^2-t)^2+4m^2(s-2m^2)+8m^4}{s^2}
\eea
This equation (\ref{schaeqn}) cannot be solved analytically and has to be computed numerically. Now, we have taken the total contribution at $\alpha_s^2$ for s-channel
annihilation which are hard scatterings without 
self energy corrections.

\end{document}